\documentclass[prl,twocolumn,showpacs]{revtex4-1}
\usepackage{graphicx}
\usepackage{times}
\usepackage{bm}
\usepackage{amsmath, verbatim}
\usepackage{amsfonts}

\def\lsim{\mathrel{\rlap{\lower4pt\hbox{\hskip1pt$\sim$}}
    \raise1pt\hbox{$<$}}}                
\def\gsim{\mathrel{\rlap{\lower4pt\hbox{\hskip1pt$\sim$}}
    \raise1pt\hbox{$>$}}}                

\def\be{\begin{equation}}
\def\ee{\end{equation}}
\def\bea{\begin{eqnarray}}
\def\eea{\end{eqnarray}}
\def\bse{\begin{subequations}}
\def\ese{\end{subequations}}

\def\be{\begin{eqnarray}}
\def\ee{\end{eqnarray}}

\begin{document}

\title{Non-locality of zero-bias anomalies in the topologically-trivial phase of Majorana wires}
\author{Tudor D. Stanescu}
\affiliation{Department of Physics and Astronomy, West Virginia University, Morgantown, WV 26506, USA}
\author{Sumanta Tewari}
\affiliation{Department of Physics and Astronomy, Clemson University, Clemson, SC 29634, USA}

\begin{abstract}
We show that the topologically trivial zero bias peak (ZBP) emerging in semiconductor Majorana wires due to soft confinement exhibits correlated
splitting oscillations as a function of the applied Zeeman field, similar to the correlated splitting of the Majorana ZBP. Also, we find that the presence of a strong impurity can effectively cut the wire in two and destroy the correlated splitting in both the trivial and the Majorana regimes. We identify a strong nonlocal effect that operates \textit{only} in the topologically trivial regime and demonstrate that the dependence of the ZBP on the confining gate potential at the {\em opposite} end in Majorana wires with two normal metal end-contacts represents a powerful tool for discriminating between topologically trivial and nontrivial ZBPs.
\end{abstract}

\maketitle

First predicted in the context of high energy physics, Majorana fermions (MF) \cite{Majorana} have come under renewed focus in low-temperature condensed matter physics \cite{Majorana,Wilczek,Frantz} as zero-energy bound states endowed with non-Abelian statistics \cite{Nayak-Wilczek}, hence, potential suitable candidates for  fault-tolerant quantum computation \cite{Kitaev-1D,Nayak-RMP}. Proposals for realizing MFs in low-temperature systems are based on fractional quantum Hall systems \cite{Nayak-Wilczek,Read-Green}, chiral $p$-wave superconductors/superfluids \cite{Read-Green,Tewari-strontium},  heterostructures of topological insulators and superconductors \cite{Fu-Kane}, and cold fermion systems \cite{Zhang-Tewari,Sato-Fujimoto}. The recently proposed scheme based on spin-orbit coupled semiconductor thin films \cite{Sau,Alicea2010,Tewari-Annals,Long-PRB} and nanowires \cite{Long-PRB,Roman,Oreg}  with Zeeman splitting and proximity induced $s$-wave superconductivity involves only conventional ingredients. The semiconductor Majorana wire -- the 1D version \cite{Long-PRB,Roman,Oreg} of the semiconductor-superconductor (SM-SC) heterostructure -- represents a direct physical realization of the one-dimensional Kitaev model \cite{Kitaev-1D}. 
The observation of a sharp zero bias conductance peak (ZBP) in charge tunneling measurement has been proposed \cite{Long-PRB,Sengupta-2001,bolech,Tewari-Lee,R1,SLDS} as a possible detection scheme for MFs localized near the ends of Majorana nanowires. The 1D SM-SC heterostructure and the associated ZBP measurements have recently attracted considerable experimental effort \cite{Mourik,Deng,Weizman,Rokhinson,Churchill,Finck,Appelbaum,Stanescu}.

Despite significant progress, the actual observation of MFs in SM nanowires is still under debate, as signatures similar to the Majorana-induced ZBPs are predicted to occur even in the topologically trivial phase in the presence of smooth confining potentials \cite{Kells}, or strong disorder \cite{Liu}. On the other hand, due to the overlap of the wave functions localized near the opposite ends,  the Majorana ZBP is characterized by splitting oscillations as a function of the Zeeman field or the chemical potential. The ZBPs measured at the opposite ends of a clean wire should be characterized by the same splitting oscillations (correlated splitting), since they involve the same wave function overlap. 
The direct observation of correlated splitting has been recently proposed \cite{Dassarma}  as a tool for identifying Majorana-induced ZBPs.

In this Rapid Communication we show that topologically trivial ZBPs emerging in SM Majorana wires due to soft confinement can also exhibit
splitting oscillations similar to those expected for the Majorana ZBP \cite{Dassarma,Rainis}. The trivial ZBPs are generated by low-energy Bogoliubov de Gennes (BdG) subgap states associated with the top occupied band and having spatial profiles characterized by maxima localized near the wire ends, similar to the Majorana wave functions. Consequently, the overlap of the wave functions localized near the opposite ends gives rise to correlated splitting oscillations. Furthermore, analogous to the behavior of the ZBPs in the Majorana regime, the presence of a strong impurity inside the superconducting segment of the wire can  destroy the correlated splitting (i.e., the splitting oscillations measured at the two ends are different). Therefore, these results show that correlated splitting oscillations, although a nontrivial quantum mechanical nonlocal effect, cannot by themselves establish the distinction between the Majorana ZBP and mundane ZBPs arising from soft confinement. Nonetheless, we demonstrate that the Majorana and the topologically trivial ZBPs depend qualitatively differently on the gate potential \textit{at the opposite end} of the wire, both in the clean case and in the presence of disorder, as long as the disorder potential does not localize the trivial low-energy states. Consequently, establishing the independence of the ZBP on the confining gate potential at the opposite end strongly indicates the Majorana nature of the ZBP in the absence of correlated splitting and completes the demonstration of Majorana modes in clean wires characterized by correlated splitting.

 Although independent splitting oscillations of the ZBPs at the two ends of a Majorana wire have been observed in recent experiments \cite{Mourik,Churchill}, correlated splitting oscillations have not yet been reported. This indicates that the superconducting segment of the wire must have strong impurity centers cutting the wire into separate pieces and destroying the correlation between the two ends. To capture this effect in both the topologically trivial and nontrivial regimes (with the ZBPs due to soft confinement and MFs, respectively) we model the most general experimental situation by considering two finite soft barrier potentials separating a central superconducting segment from the normal end segments  and a strong impurity site inside the superconducting region. The normal end segments of the semiconductor wire are coupled to metallic contacts. In this geometry, in addition to the correlated splitting (for weak or no impurity), we discover a second nonlocal effect -- a nonlocal dependence of the ZBP at one end on the barrier height at the other end -- that occurs only in the topologically trivial regime. We show that in the topologically trivial regime the ZBP can weaken and disappear if the height of the barrier potential at the opposite end is lowered below a certain value. This effect occurs only in the topologically trivial regime because the subgap BdG states responsible for the ZBP are significantly more extended than the Majorana modes and can leak out of the SC region into the metallic contacts, if the height of the confining barrier potential is decreased. By contrast, such a nonlocal effect is absent in the Majorana regime, since the Majorana wave function is tightly bound to the SC-normal metal interface and remains unaffected by the confining potential at the opposite end. Our theory predicts that, although the correlated splitting oscillations as a function of the Zeeman field and/or the chemical potential cannot by themselves distinguish between the Majorana ZBP and trivial ZBPs due to soft confinement, this second nonlocal effect -- the dependence of the ZBP on the gate potential at the opposite end -- provides the tool required for discriminating between these possible sources of zero bias anomalies.

The model used in the calculations consists of a rectangular semiconductor (SM) nanowire with dimensions  $L_x\gg L_y \gg L_z$ described in the limit $L_x\rightarrow \infty$ by the effective tight-binding BdG Hamiltonian
\begin{eqnarray}
H_{nm}(k) &=&[\epsilon_{nm}(k) -\mu \delta_{nm}]\tau_z + \Gamma\delta_{nm}\sigma_x\tau_z \nonumber \\
&+& \alpha k \delta_{nm} \sigma_y\tau_z - i \alpha_y q_{nm} \sigma_x +\Delta_{nm}\sigma_y\tau_y,      \label{Hnm}
\end{eqnarray}
where $k$ is the wave number, while $\sigma_i$ and $\tau_i$ are Pauli matrices in the spin and particle-hole (p--h) spaces, respectively.  In Eq.~(\ref{Hnm}) $n =(n_y,n_z)$ and  $m=(m_y,m_z)$ are quantum numbers for the confinement-induced bands described by the transverse wave functions $\phi_n(y, z) \propto \sin( n_y\pi y/L_y)\sin( n_z\pi z/L_z)$, $\epsilon_{nm}$ are the proximity-renormalized energies of the SM wire without spin--orbit coupling, $\alpha=a\alpha_y=0.2$ eV\AA ~(with $a\approx 6.5$\AA~ being the lattice constant) is the Rashba coupling, $\Gamma= g^* \mu_B B/2$ (with $\mu_B$ the Bohr magneton and $g^*=50$) is the Zeeman field applied along the $x$ direction,  $\mu$ is the chemical potential, $q_{nm}=-q_{mn}$ are interband spin-orbit coupling matrix elements, and $\Delta_{nm}$ is the proximity-induced pair potential. The 
parameters $\epsilon_{nm}$, $q_{nm}$, $\Delta_{nm}$ are determined numerically following  Ref.~[\onlinecite{SLDS}]. Note that below we use the term ``band" for a pair of spin subbands that are degenerate for $\Gamma=0$ and $\alpha=0$. Also, we use the notation $\Delta\mu$ for the chemical potential measured relative to the energy of the top occupied band at $k=0$ and $\Gamma=0$. 
The condition $L_y \gg L_z$ was imposed to avoid additional complications arising from the dependence of the SC proximity effect on the wire thickness \cite{Stanescu-Dassarma}. In this limit, the relevant confinement-induced bands are characterized by $n_z=1$ and the induced pair potential becomes $\Delta_{nm} = \gamma \Delta_0/(\gamma+ \Delta_0) \delta_{nm}$, where $\Delta_0$ is the bulk SC gap and $\gamma$ is the effective semiconductor-superconductor coupling \cite{Stanescu-Dassarma}. In the calculations we consider finite wires with a position-dependent coupling $\gamma(x)$ that is nonzero along a central section separated from the normal end sections by a local gate potential $V(x)$, which also includes the impurity contribution [see Fig. \ref{Fig2}(a)].

\begin{figure}[tbp]
\begin{center}
\includegraphics[width=0.48\textwidth]{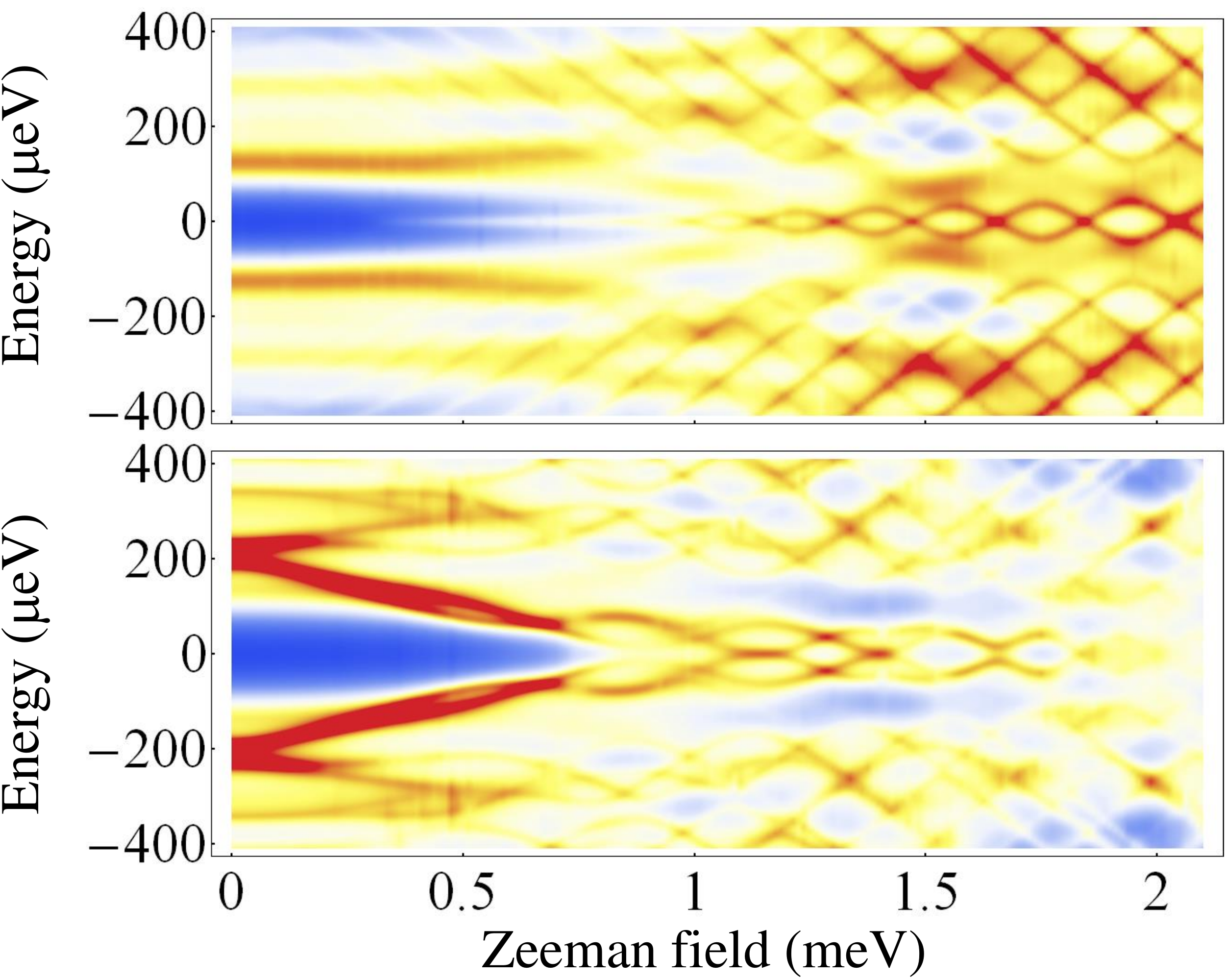}
\vspace{-7mm}
\end{center}
\caption{(Color online) Splitting oscillations of the zero-bias anomalies in the Majorana-supporting phase (top panel) and the topologically trivial phase (bottom). The local density of states integrated over the left barrier region is shown as a function of the Zeeman field for a $3.6 \mu$m nanowire proximity coupled to a SC and two normal metal end contacts. The superconducting central region ($1.8 \mu$m) is separated from the normal regions by two potential barriers (see Fig. \ref{Fig2}). Identical splitting patterns characterize the right-end LDOS. Note the absence of a gap-closing signature in the Majorana regime (top).  By contrast, the topologically trivial regime (bottom) is characterized by strong low-field features associated with the states  responsible for the zero bias anomaly.}
\vspace{-6mm}
\label{Fig1}
\end{figure}

Increasing the Zeeman splitting drives the system through a topological quantum phase transition (TQPT) at $\Gamma=\Gamma_c(\mu)$ into a topological phase characterized by  zero-energy Majorana bound states localized near the wire ends. The bulk quasiparticle gap necessarily vanishes at the TQPT \cite{Read-Green,Long-PRB,SLDS}, but 
this closing of the gap, while visible in the total density of states, may have no signature in the local density of states (LDOS) at the end of the wire \cite{Stanescu-Tewari}. In the topologically-trivial phase characterized by $\Gamma < \Gamma_c(\mu)$ the system has no Majorana bound states, but low-energy states with significant spectral weight at the wire ends are still possible in the presence of a soft confinement potential \cite{Kells,Stanescu-Tewari-2,Roy-Tewari}. 
To compare the low-energy physics in the topologically-trivial and Majorana regimes, we consider a quasi-1D nanowire with four occupied bands and two values of the chemical potential, one near the minimum of the top band ($\Delta \mu$ small) and another that cuts both subbands of the top-most band ($\Delta \mu$ large).

\begin{figure}[tbp]
\begin{center}
\includegraphics[width=0.48\textwidth]{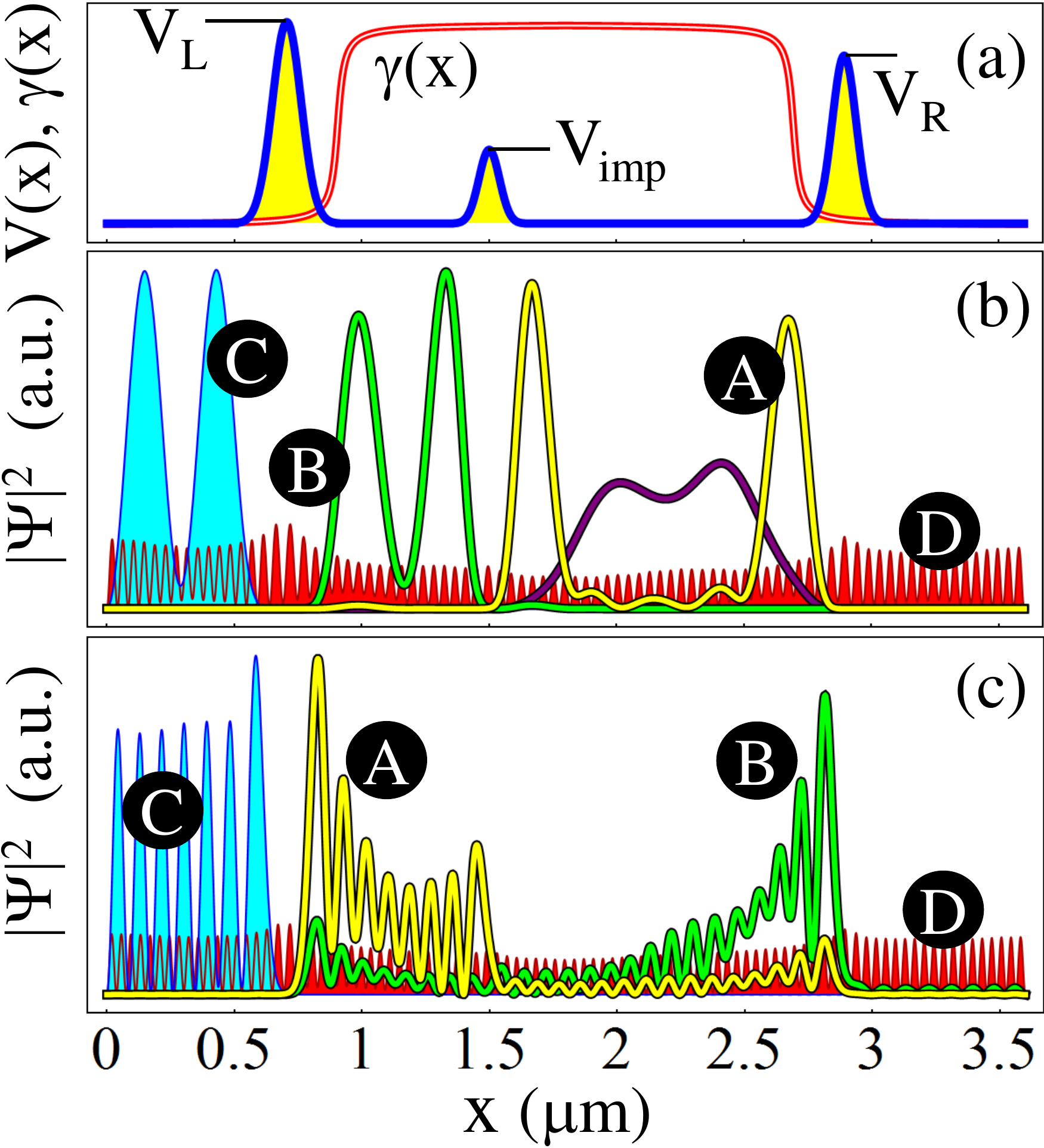}
\vspace{-8mm}
\end{center}
\caption{(Color online)  (a) Position-dependent effective SM-SC coupling, $\gamma(x)$, and local potential $V(x)$. The left and right barrier heights are $V_L$ and $V_R$, respectively, and $V_{\rm imp}$ is the amplitude of a strong impurity potential.  (b) Typical low-energy states in the Majorana regime characterized by a chemical potential near the bottom of the fourth band, $\Delta\mu=0$, and a Zeeman field $\Gamma=0.45$mev. (c) Low-energy states in the topologically-trivial regime with $\Delta\mu=3$meV and $\Gamma=0.8$meV. States A and B correspond to the top (fourth) occupied band and are responsible for the ZBPs, states C are top band states localized inside the normal regions, while D are states associated with the lower energy bands and extend throughout the entire wire. The purple (dark gray) line in panel (b) shows a finite-energy bulk-type state from the Majorana band. Note that, because of its low amplitude inside the barrier region, this type of state is not ``visible'' in local measurements near the ends of the SC region.  States that penetrate into the normal regions (C and D) hybridize strongly with the normal metal continuum and generate very broad spectral features. Parameters: $V_L=6$meV, $V_R=5$meV, $V_{\rm imp}=2$meV, $\gamma_{\rm bulk}=0.3$meV, and $\Delta_0=1.5$meV (bulk SC pair potential).}
\vspace{-6mm}
\label{Fig2}
\end{figure}

In Fig.~\ref{Fig1} we show the splitting oscillations of the zero bias peak at the left end of the SC segment of a clean wire as a function of the Zeeman field in both the Majorana-supporting phase ($\Gamma > \Gamma_c$, $\Delta \mu$ small) and the topologically trivial phase ($\Gamma < \Gamma_c$, $\Delta \mu$ large). We find that identical oscillation patterns are present at the right end of the wire, a clear signature of correlated splitting. In both cases the states responsible for the ZBP (the Majorana state in the topological regime and two low-energy in-gap states in the topologically trivial regime) are BdG states associated with the top occupied band and characterized by maxima near the wire ends and by decay lengths that depend strongly on $\Delta\mu$ and (weakly) on $\Gamma$. 
The overlap of the wave functions localized near the wire ends generates the splitting of the ZBP. In a clean system, since the same wave function overlap determines the zero energy splitting at both ends, the splitting oscillations are correlated.

\begin{figure}[tbp]
\begin{center}
\includegraphics[width=0.48\textwidth]{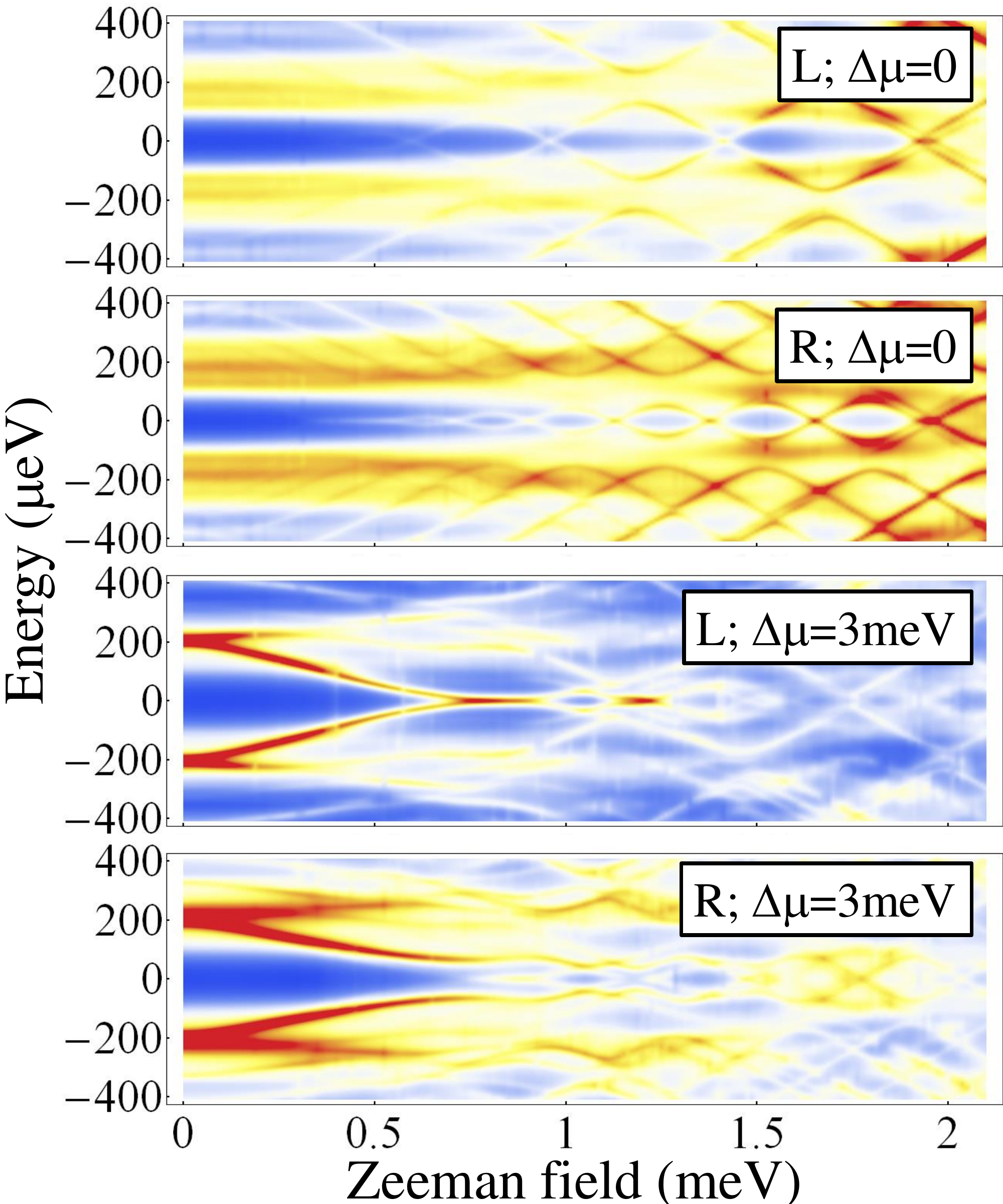}
\vspace{-8mm}
\end{center}
\caption{(Color online) Uncorrelated splitting in a wire with a strong impurity ($V_{\rm imp}=2$meV) in the Majorana ($\Delta\mu=0$) and the topologically trivial ($\Delta\mu=3$meV) regimes. The parameters of the system are the same as in Fig. [\ref{Fig2}]. The Majorana splitting oscillations measured at the left and right end of the SC wire segment are generated by the B and A modes, respectively (see Fig. [2]) and have different amplitudes because the impurity separates the wire into two segments of different length.  In the trivial regime, the ZBP at $\Gamma<1$meV is produced by the A mode, which, for $\Gamma<0.75$meV,  has a very small amplitude at the right end and does not generate a visible signature (while mode B remains gapped).}
\vspace{-4mm}
\label{Fig4}
\end{figure}

Next, since correlated splitting of the ZBPs has not yet been observed in the experiments, we consider the more general situation that includes soft barrier potentials near the ends of the SC wire segment and a strong impurity potential asymmetrically placed inside the SC segment (Fig.~\ref{Fig2}). If the impurity potential is strong enough,  the splitting oscillations as a function of the Zeeman field become uncorrelated in both the Majorana and the topologically trivial regimes, as shown in Fig. \ref{Fig4}. This absence of correlation between the ZBPs measured at the two ends is caused by the strong impurity potential, which effectively cuts the SC wire into two segments of unequal spatial extent if $V_{\rm imp}\gg\Delta\mu$, i.e., in the Majorana regime. 
The low-energy states responsible for the ZBPs visible at the two ends of the SC wire  are shown in Fig. \ref{Fig2} (states A and B). Note that, in the Majorana regime, the states A and B are confined insides the two SC segments separated by the impurity, while in the topologically trivial regime the two states penetrate through the impurity potential and can still create correlated features in the LDOS measured at the opposite ends, although, as evident from the lower panels of Fig. \ref{Fig4}, these features may be difficult to identify.  Another striking difference between the Majorana and the topologically-trivial regimes is represented by the so-called ``gap-closing'' signature \cite{Stanescu-Tewari,Stanescu-Tewari-2}. Note that in the Majorana regime the emergence of a ZPB is not accompanied by any visible signature associated with the closing of the quasiparticle gap at the TQPT, as shown in both Figs. \ref{Fig1} and \ref{Fig4}. This is due to the low amplitude near the wire ends of the relevant bulk-type states [see Fig. \ref{Fig2}(b)]. The Majorana (i.e. lowest energy) mode itself is characterized by a similar profile at low values of the Zeeman field, and it becomes ``invisible'' to local end-of-wire probes for $\Gamma<\Gamma_c$.  By contrast, the states responsible for the ZBP in the topologically-trivial phase have a profile that does not change qualitatively with $\Gamma$ and, consequently, can be traced as a function of the Zeeman field starting at $\Gamma=0$ (see Figs. \ref{Fig1} and \ref{Fig4}). Note, however, the possibility that two low-energy states characterized by different splitting patterns [e.g., A and B shown in Fig. \ref{Fig2}(c)] contribute to the low-energy end-of-wire LDOS. As shown in Fig. \ref{Fig4} (bottom panel),  a ZBP is visible at the right end of the nanowire for $\Gamma<1$ meV that appears to have been generated without a complete closing of the gap. This peak is produced by the A mode [Fig. \ref{Fig2}(c)], which, for $\Gamma<0.75$ meV,  has a very small amplitude at the right end and does not generate a visible signature in the LDOS. Such a ZBP in the topologically trivial regime, which does not appear to be accompanied by a full ``gap-closing" signature \cite{Stanescu-Tewari,Stanescu-Tewari-2}, has striking similarities with the Majorana ZBP. However, unlike the absence of a gap-closing signature in the Majorana regime, this feature is not robust against variations of the confining potential amplitude.  Below, we discuss in more detail the nonlocal effect due to barrier potentials and demonstrate that it can help discriminate between the two types of ZBPs in semiconductor Majorana wires.

\begin{figure}[tbp]
\begin{center}
\includegraphics[width=0.48\textwidth]{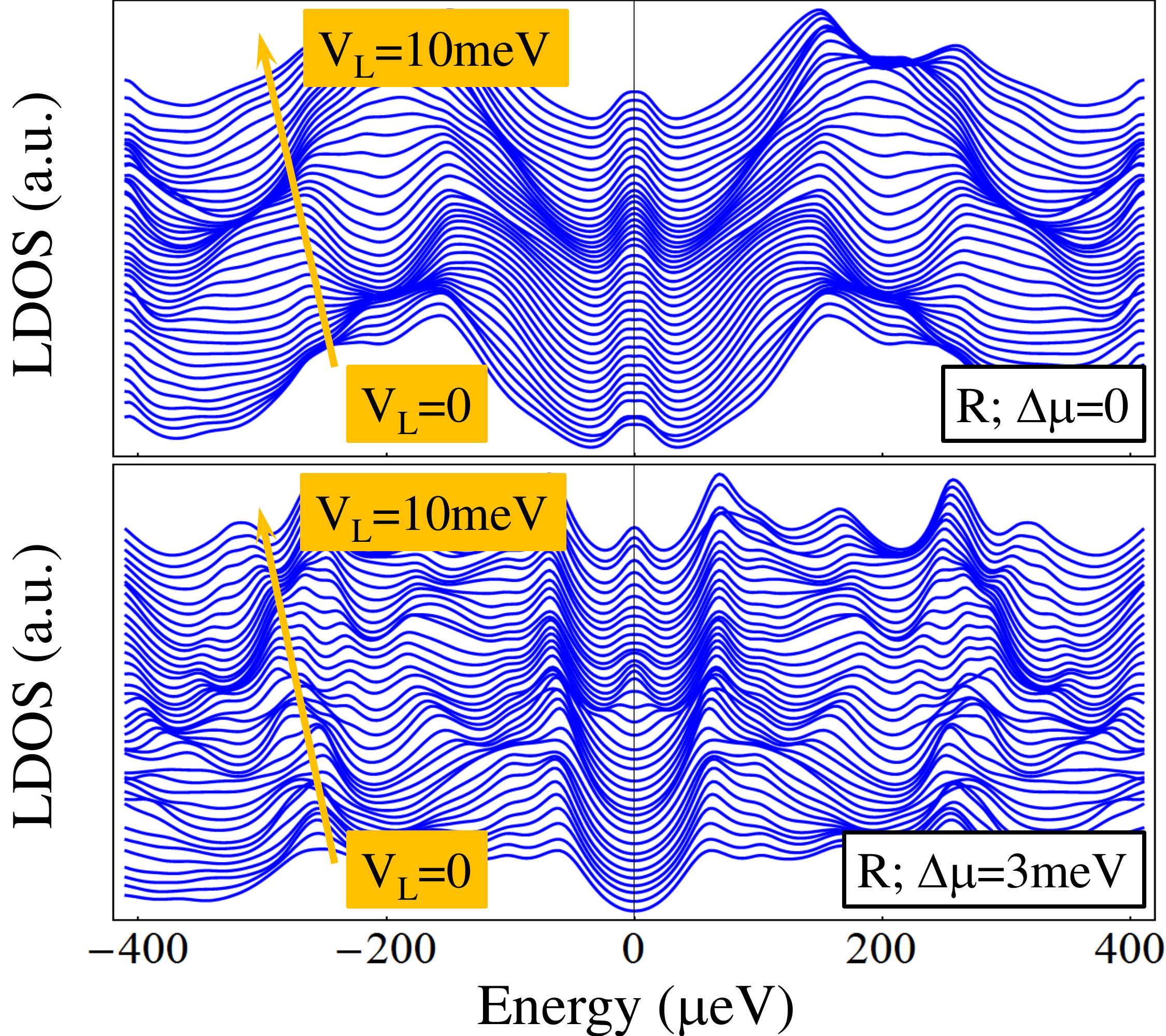}
\vspace{-6mm}
\end{center}
\caption{(Color online) Nonlocal effect in the topologically trivial phase. The dependence of the right end ZBP on the barrier hight at the left end is shown for the Majorana (top) and the topologically-trivial (bottom) regimes. The other parameters are the same as in Fig. \ref{Fig2}. Note that the Majorana peak is weakly affected by the potential barrier at the opposite end, while the trivial ZBP disappears when the barrier at the opposite end is lower that a certain value as mode A (see Fig. \ref{Fig2}) leaks into the normal region.}
\vspace{-8mm}
\label{Fig5}
\end{figure}

The dependence of the LDOS at the right end of the SC wire on the barrier height $V_L$ at the left end is shown in Fig. \ref{Fig5}. In the Majorana regime (top panel), the LDOS is characterized by a ZBP that is not affected by the confining potential at the opposite end. This behavior is due to the localized nature of the Majorana bound state and does not depend on the presence of impurities. In essence, the Majorana bound state responsible for the ZBP will maintain its spatial profile regardless of the fate of its counterpart, which could remain localized neat the impurity or leak into the normal region for low values of $V_L$ in clean enough wires. By contrast, in the topologically trivial regime the ZBP at the right end disappears if the height of the barrier potential \textit{on the other end} is lowered below a certain value. This effect occurs because the subgap BdG states responsible for the ZBP in the topologically-trivial regime are characterized by Fermi momenta significantly larger than those of the  Majorana mode. As a result, these states can penetrate the finite potential barriers and leak into the metallic contacts. As we discussed before \cite{Stanescu-Tewari-2}, such leaking into the contacts and the corresponding hybridization with the metallic states result in broad subgap features responsible for the soft gap seen experimentally \cite{Mourik}. Only states confined within the SC segment can generate sharp features in the LDOS. For state A in Fig.~\ref{Fig2}(c) lowering  the barrier potential at the left end generates deconfinement and hybridization with the metallic states and, consequently, the disappearance of the ZBP at the right end. State B, on the other hand, has finite energy in the relevant field range,  $0.75$ meV$<\Gamma<1$ meV. Note that the nonlocal effect described here is distinct from the nonlocal current correlations discussed recently \cite{Liu-Law}. Also, the strong dependence of the ZBP on the pinch-off gate at the other end described here is a characteristic of the topologically-trivial phase with $\Delta\mu\gg \Delta_{nn}$ in the presence of weak disorder and soft confinement, rather than a signature of a disorder-driven TQPT \cite{Fregoso} in systems with $\Delta\mu\approx 0$.  

In conclusion, we have identified a nonlocal effect -- the dependence of the ZPB at one end of the wire on the confining potential at the opposite end -- that emerges only in the topologically-trivial phase of a Majorana wire and can be used as a powerful tool for discriminating between topologically-trivial and Majorana-induced ZBPs. To practically  demonstrate zero-energy MFs, we propose the following three-step protocol: 1)  Establish the emergence of a ZBP at a finite Zeeman field. If the ZBP is preceded by a gap-closing signature, stop. 2) Establish the presence of correlated ZBP splitting oscillations by performing a two-end tunneling experiment on a relatively short wire ($L_x\sim 1\mu$m). This ensures that the wire is clean enough, so that only two spatially-overlapping low-energy states are responsible for the ZBPs. 3)  Determine the dependence of the ZBP measured at one end of the wire on the confining potential at the other end. If the ZBP is robust, it is generated by a zero-energy Majorana bound state. 


{\em Acknowledgment:} This work was supported by WV HEPC/dsr.12.29, NSF (PHY-1104527), and AFOSR (FA9550-13-1-0045)



\end{document}